\documentclass[11pt]{article}
\pdfoutput=1  
\usepackage{jcappub}
\usepackage{graphicx}
\usepackage{amsmath,amssymb}
\graphicspath{{./figures/}}
\usepackage{appendix}
\usepackage{placeins}
\usepackage[normalem]{ulem}
\usepackage[utf8]{inputenc}

\input epsf
\usepackage{mathtools}
\usepackage{amsfonts}
\usepackage{upgreek}
\usepackage{latexsym}
\usepackage{stfloats}
\usepackage{afterpage}

\newcommand{\be}{\begin{equation}}
\newcommand{\ee}{\end{equation}}
\newcommand{\beq}{\begin{equation}}
\newcommand{\eeq}{\end{equation}}
\newcommand{\bea}{\begin{eqnarray}}
\newcommand{\eea}{\end{eqnarray}}

\newcommand{\bk}{{\mathbf k}}

\newcommand{\HH}{{\cal H}}

\newcommand{\De}{\Delta}

\newcommand{\La}{\Lambda}
\newcommand{\si}{\sigma}

\newcommand{\Om}{\Omega}

\newcommand{\pa}{\parallel}

 \definecolor{magenta}{rgb}{0.1,0.98,0.6}

\definecolor{dgreen}{rgb}{0,0.6,0.0}

\definecolor{cyan}{rgb}{0,0.6,0.6}

\title{Small scale effects in the observable power spectrum at large angular scales}
\author{William L. Matthewson}
\author{and Ruth Durrer}
\affiliation{Universit\'e de Gen\`eve, D\'epartement de Physique Th\'eorique and Centre for Astroparticle Physics,
24 quai Ernest-Ansermet, CH-1211 Gen\`eve 4, Switzerland}

\emailAdd{william.matthewson@unige.ch}
\emailAdd{ruth.durrer@unige.ch}

\abstract{In this paper we show how effects from small scales can enter the angular-redshift power spectrum $C_\ell(z,z')$. In particular, we show that spectroscopic surveys with high redshift resolution are already affected on large angular scales, i.e. at low multipoles, by features from small scales.  When considering the angular power spectrum with spectroscopic redshift resolution, it is therefore important to account for non-linearities relevant on small scales, even at low multipoles. This may also motivate the use of the correlation function in relatively wide redshift bins, which is not affected by non-linearities on large scales, instead of the angular power spectrum. The extent to which small-scale effects become visible on large scales, which is more relevant for bin auto-correlations than for cross-correlations, is quantified in detail.}

\begin{document}

\maketitle

\section{Introduction}\label{s:intro}
In  cosmology, positions of galaxies are  truly observed as redshifts and angles. If we want to determine the correlation properties of galaxies (or classes of galaxies) in a model-independent way, we must study the redshift-dependent angular power spectra, $C_\ell(z,z')$, or the angular correlation functions, $\xi(\theta,z,z')$. Assuming statistical isotropy, these 2-point functions fully determine the 2-point statistics of the galaxy distribution and, with the additional assumption of Gaussianity, they determine all statistical quantities.  Whenever  redshifts and angles are converted into length scales,  assumptions about the background cosmology are made. At very low redshift, $z\ll1$, the distance is entirely determined by the Hubble parameter, $r(z)\simeq z/H_0$  and the model dependence is encoded in the 'cosmological unit of length' given by $h^{-1}$Mpc, where $h=H_0/100$km/sec/Mpc. However, at redshifts of order unity and larger, the full cosmological model enters in the determination of $r(z)$.
This fact has prompted a tendancy in the field to prefer the directly-observable angular-redshift power spectra and correlation function. We cite Refs.~\cite{Bonvin:2011bg,Challinor:2011bk,Jeong:2011as,DiDio:2013bqa,DiDio:2013sea,Alonso:2015uua,Lorenz:2017iez} as examples.

In this paper we study the following question: When analyzing a spectroscopic dataset, in which redshifts are very well known, that is $\si_z \lesssim 10^{-3}$, does the precise radial information about the galaxy position affect the power spectrum at low multipoles of $\ell\lesssim 100$?

We shall see that the answer to this question is yes, as already noted in~\cite{Jalilvand:2019brk}. Here we perform a detailed study of the amplitude of the effect and its origin. We find that small scale effects, including changes in the small-scale power due to non-linearities especially, significantly affect all spectroscopic $C_\ell(z,z)$'s including, most interestingly, at low $\ell$. 
This finding is not so surprising and is actually just another way to understand why the Limber approximation~\cite{Limber:1954zz,LoVerde:2008re}, which selects a single  scale  $k$ relevant for a given $\ell$, totally fails for spectroscopic number count $C_\ell(z,z)$'s, see e.g.~\cite{Assassi:2017lea,Matthewson:2020rdt}.
We study how the contributions of these effects on large scales decay if we either smear out the $C_\ell(z,z)$'s over a sufficiently large redshift window or consider cross-correlations, $C_\ell(z,z')$ with sufficiently large $|z-z'|$. Contrary to correlations of galaxy number counts alone, lensing-lensing or lensing-number counts cross-correlations (galaxy-galaxy-lensing) are insensitive to the appearance of small scale effects at low multipoles, due to the broad kernel of the shear and magnification integrals. 

This paper is organized as follows: in the next section we study the origin of the imprint of small scale contributions on the $C_\ell(z,z)$'s from galaxy number counts at low $\ell$, in general. In Section~\ref{s:nlin} we investigate the effect induced at low $\ell$'s by a change in shape in the high $k$ power spectrum, due to non-linearities for example, as a function of redshift as well as window width. Our aim here is an estimate of the amount by which the $C_\ell$'s are affected, but not a detailed non-linear modelling of them. The latter could be achieved by high resolution N-body simulations and goes beyond the scope of this work. In Section~\ref{s:cross} we investigate cross-correlations and in Section~\ref{s:con} we end with our conclusions.

\section{Small scale contributions to low \texorpdfstring{$\ell$}~~spectroscopic \texorpdfstring{$C_\ell(z,z)$'s} 
~~: Generics} \label{s:small} 
If we consider correlations of galaxies at fixed redshift with small redshift uncertainty $\si_z$, their comoving radial separation is smaller than
\be\label{e:rk}
r(z,\si_z) = \frac{\si_z}{H(z)}\,,  \quad \mbox{corresponding to a radial wave number} \quad k_{\pa}(z,\si_z)  = \frac{2\pi H(z)}{\si_z}  \,.
\ee
Typical spectroscopic surveys have redshift resolution of $\si_z\sim 10^{-3}$ or better.
As an example, $r(1,10^{-3}) =1.7h^{-1}$Mpc, which is well within the non-linear regime at $z=1$.

To find a quantitative estimate, we consider the flat sky approximation for $C_\ell(z,z')$, which is excellent for very close redshifts~\cite{Matthewson:2020rdt}.
\be
C_\ell(z,z') = \frac{1}{2\pi r^2(\bar z)}\int_{-\infty}^{\infty}dk_{\pa} P_g(k,\bar z)e^{-ik_{\pa}(z-z')/H(\bar z)} \,,
\ee
where $\bar z=(z+z')/2$ is the mean redshift, $k =\sqrt{k_{\pa}^2 +\ell^2/r^2(\bar z)}$ and $P_g = b^2(z)P_m$ denotes the galaxy power spectrum. $P_m$ is the matter power spectrum and $b(z)$ is the galaxy bias (we neglect non-linear bias).
Integrating this over $z$ and $z'$, with a tophat window of width $\si_z$ centered at $\bar z$, we obtain
\be\label{e:flatP}
C_\ell(\bar z,\si_z) = \frac{1}{2\pi r^2(\bar z)}\int_{-\infty}^{\infty}dk_{\pa} P_g(k,\bar z)j_0^2\left(\frac{k_{\pa}\si_z}{2H(\bar z)}\right) \,.
\ee
Strictly speaking, the above expression is for the density term only. In order to include redshift space distortions (RSD), we have to replace the power spectrum by 
\be\label{e:DRSD-lin}
P_{D+RSD}(k,\mu,\bar z)= (b(z)+f(z)\mu^2)^2P_m(k,\bar z)\,,
\ee
where $\mu=k_\pa/k$ is the direction cosine of $\bk$ with the forward direction and $f(z)$ is the growth rate, see e.g.~\cite{Durrer:2020fza}. This is the power spectrum from linear perturbation theory which we use to determine the so-called `standard terms' which comprise density and RSD correlations and which are used in the present analysis.

We have compared approximation \eqref{e:flatP} (replacing $P_g$ by $P_{D+RSD}$) with the {\sc class}-code~\cite{Blas:2011rf} and found that the difference is smaller  than 1\% for windows of size $\si_z< 0.01$. For larger windows the error slowly grows  and reaches about 4\% for $\sigma_z=0.1$ and $\ell=100$.

\begin{figure}[!ht]
\centerline{\includegraphics[width=0.8\linewidth]{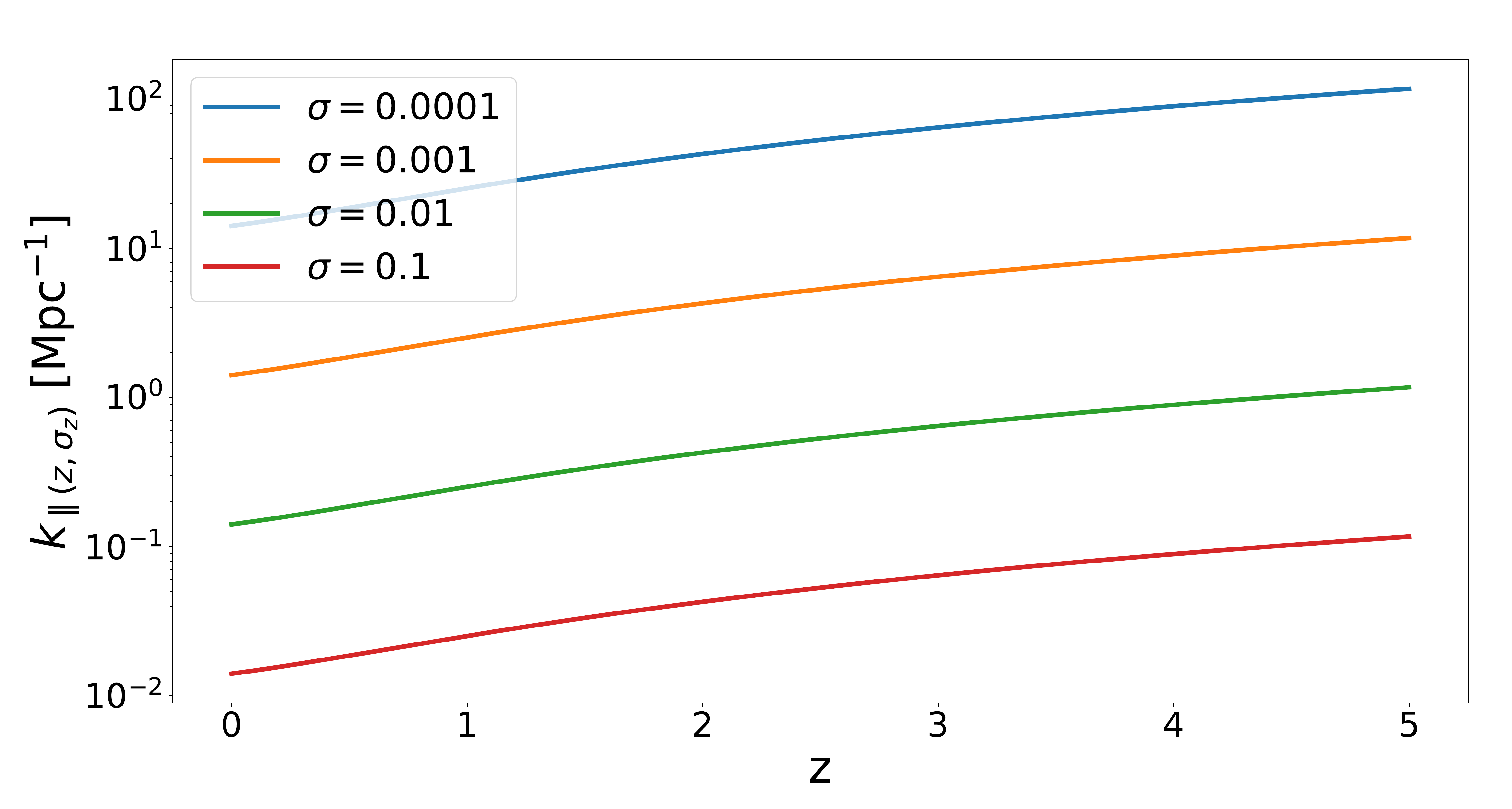}}
\caption{\label{f:kzsi}  We show $k_\pa(z,\si_z)$ as a function of redshift, with different colours indicating different values of $\sigma_z$ ranging from $10^{-4}$ to $10^{-1}$. As is clear from the second part of (\ref{e:rk}), for wider redshift windows the most important contributions to the spectrum come from smaller value of $k_{\parallel}$.}
\end{figure}

The spherical Bessel function, $j_0$,  acts as a `low-pass filter', meaning that only modes with $k_{\pa} \leq 2\pi H(\bar z)/\si_z= : k_\pa(z,\si_z)$ significantly contribute to the angular power spectrum. The smaller $\si_z$, the higher the values of $k_{\pa}$ which contribute.
In Fig.~\ref{f:kzsi} we plot $k_\pa(z,\si_z)$ as a function of redshift for some values of $\si_z$.

This also explains why the Limber approximation fails for narrow redshift bins. The Limber approximation, which replaces $k=\sqrt{k_\pa^2+\ell^2/r(z)^2}$ by $(\ell+1/2)/r(z)$, is  a reasonable approximation only for values of $\ell$ with  $k_\pa(z,\si_z)\lesssim \ell/r(z)$.  Inserting $k_\pa(z,\si_z)$, we find that this requires $\ell>\ell(z,\si_z)$, with
\be
\ell(z,\si_z) = r(z)k_\pa(z,\si_z)=2\pi H(\bar z)r(z)/\si_z= 2\pi \int_0^zdz'\frac{H(z)}{H(z')} > 2\pi\frac{z}{\si_z} \,.
\ee
For $z\sim 1$ and $\si_z\lesssim 10^{-3}$, this means $\ell(z,\si_z)>6000$. Thus, the Limber approximation is completely out of the question at the values of $\ell$ that we consider here and, as we shall later see, non-linearities enter the  spectrum $C_{\ell}(z)$ for all values of $\ell$.

While those at equal redshift considered here are calculated using the flat sky approximation, our later numerical results at unequal redshift are obtained with the code `CAMB sources'~\cite{Challinor:2011bk} which allows for a simpler adjustment of the $k_{\max}$ in the integration of the spectra than {\sc class}. For better stability in the case of unequal redshifts, we use Gaussian windows with full width at half maximum given by $\si_z$ throughout.

Of course, the extent to which the contributions from high $k_\pa$ are relevant also depends on the shape of the power spectrum. For example, we expect them to be more relevant for the non-linear power spectrum which is boosted, especially at small scales. In this section we calculate the linear angular power spectrum, as well as the non-linear halofit ~\cite{Takahashi:2012em} spectrum as an example of a different spectrum shape, taking cut-offs at different values $k_{\max}$ in the integral over $k$. In particular, we consider the spectroscopic case of a narrow window width $\sigma_z$ that modifies the relevant value of $k_\parallel$, see Fig~\ref{f:kzsi}. In this way we can investigate how sensitive the resulting low $\ell$ portion of the spectrum is to high $k$ contributions, and whether the presence of extra power at high $k$ in the case of the halofit spectrum makes a significant difference. We are not concerned here with the precise shape of the {\em true} non-linear power spectrum, but rather whether a sizable change in power at small scales makes a significant change to the amplitude of the angular power spectrum, even at low $\ell$.

\begin{figure}[!ht]
\includegraphics[width=0.9\linewidth]{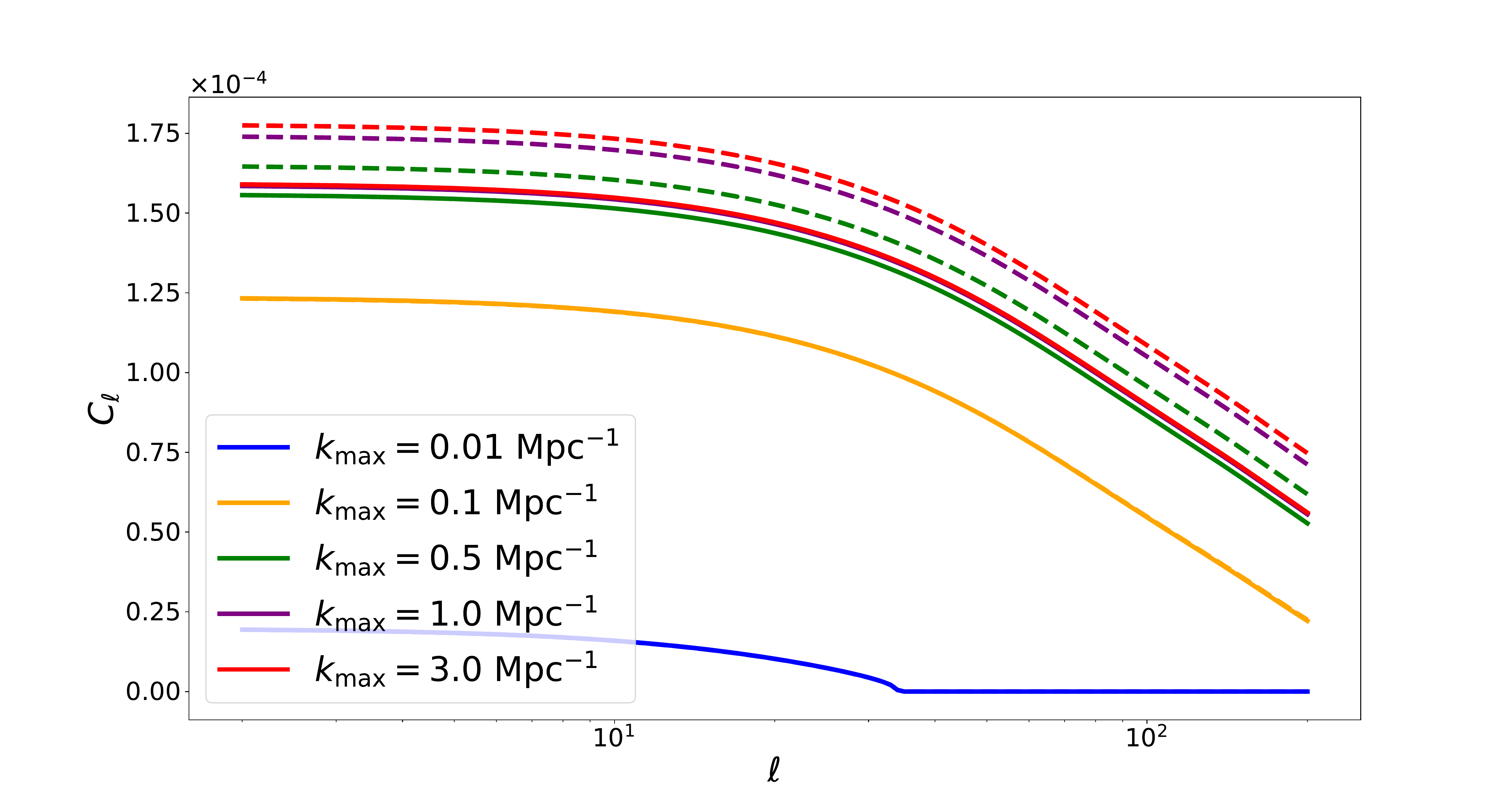}
\caption{\label{f:Clk2a}  We show 
$C_\ell(\bar z=1,\si_z=10^{-3})$ computed using various values of $k_{\rm max}$, shown in different colours. The solid lines represent the linear spectra, while the dashed lines represent the halofit corrections. As $k_{\rm max}$ increases, the spectra converge to their accurate values at each $\ell$. For the lowest values of $k_{\rm max}$, the spectrum at higher $\ell$'s does not obtain any contributions from the range of $k$ considered, and drops to zero. The difference between the linear spectrum and halofit increases with $k_{\max}$. The spectra for $k_{\max}=1/$Mpc and $k_{\max}=3/$Mpc are very close, which indicates that the spectrum is converging at $k_{\max}\simeq 3/$Mpc.}
\end{figure}

\begin{figure}[!ht]
\includegraphics[width=0.9\linewidth]{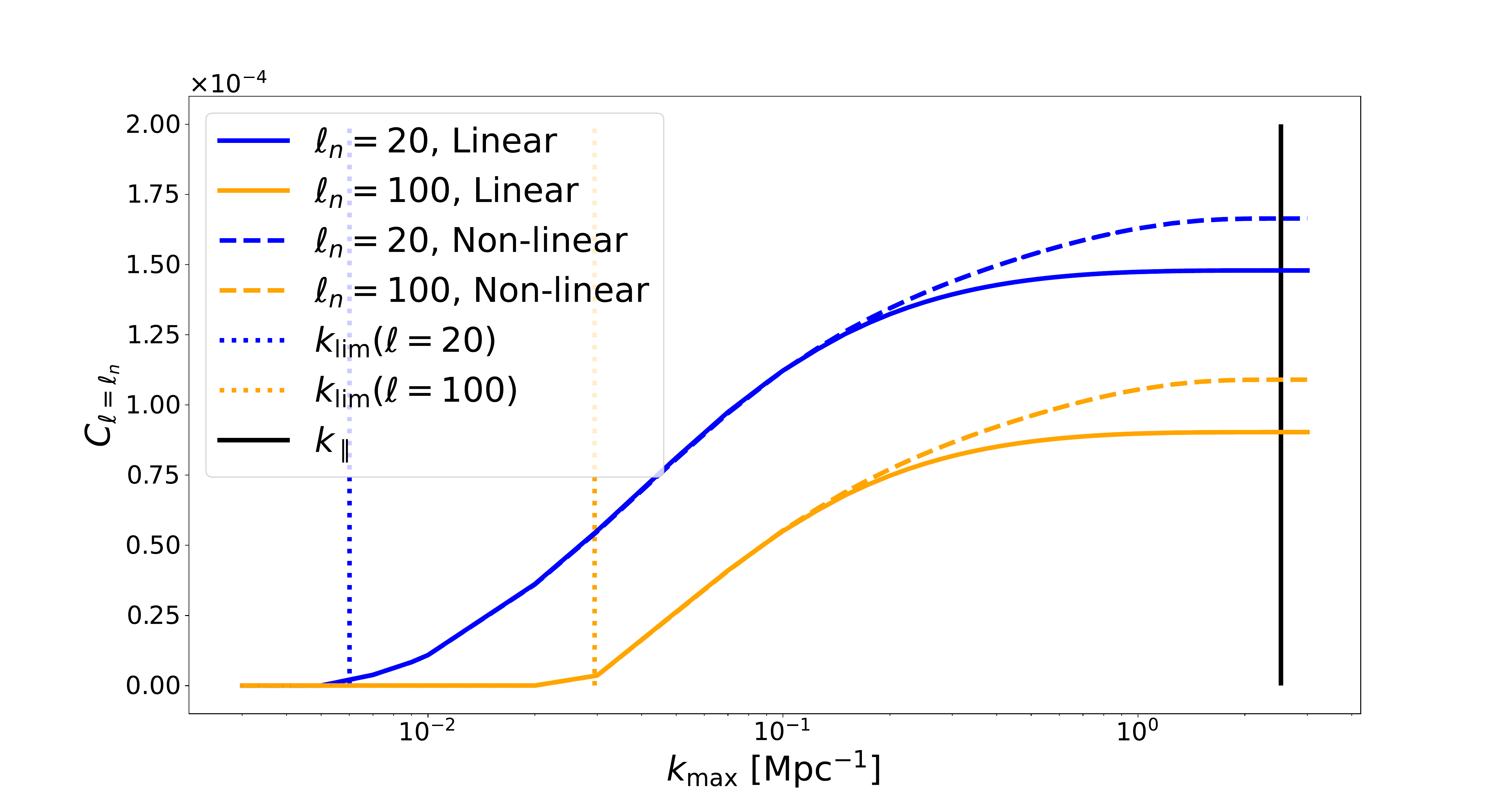}
\caption{\label{f:Clk2b}
We show $C_{20}(\bar z=1,\si_z=10^{-3})$ (blue)  and $C_{100}(\bar z=1,\si_z=10^{-3})$ (orange) from the linear (solid lines) and halofit (dashed lines) power spectra as a functions of $k_{\rm max}$, the upper boundary of the $k$-integral. The values of $k_\parallel(z,\si_z)$ and $k_{\rm Lim}(\ell,z)$ are also indicated. From this plot we see that  $k_{\max}\simeq k_\parallel(z,\si_z)$ is required for the halofit spectra to  converge. }
\end{figure}

As an example, in Fig.~\ref{f:Clk2a} we plot $C_\ell(\bar z=1,\si_z=10^{-3})$, integrating up to different values of $k_{\max}$. The power missing for small $k_{\max}$ at low $\ell$ is as significant as at high $\ell$. Hence the value of $\ell$ is not simply related to the value of $k$ which gives the dominant contribution to the angular power spectrum. For the lowest  $k_{\max}$ there is no difference between the linear power spectrum and halofit, but for  $k_{\max}\gtrsim 0.1$ the halofit power spectrum boosts all $C_\ell$'s. However, while for $k_{\max}=1$ and $k_{\max}=3$ the linear power spectra overlap, the non-linear spectra still differ by about 2\%. This shows that the effect of the non-linearities on the shape also affects the angular power spectrum at low $\ell$.

In Fig.~\ref{f:Clk2b} we plot $C_{20}$ (blue) and $C_{100}$ (orange) for $\bar z=1,\si_z=10^{-3}$
as functions of $k_{\max}$. As vertical lines we also indicate $k_\pa(z,\si_z)$ (black) and $k_{\rm Lim}(z,\ell) = (\ell+1/2)/r(z)$ (dotted, blue and orange respectively). The value $k_\pa$ is roughly where   convergence of the spectra is reached\footnote{Even though $k_{\max}$ is the value of $k$ not $k_\pa$, this difference is irrelevant for $k_\pa\gg k_{\rm Lim}$, since in this regime $k\simeq k_{\pa}(1+\frac{1}{2}(k_{\rm Lim}/k_\pa)^2)$.}. It is evident that non-linearities have a significant effect, indicated by the dashed lines.

For comparison, at $z=1$ we obtain $k_{\rm Lim}\simeq 4.36\times 10^{-4}(\ell+1/2)h/$Mpc and $k_\pa(z,\si_z = 10^{-3}) \simeq 3.75h/$Mpc. We use $\La$CDM with $\Om_m=0.31$, $\Om_\La=0.69$ and $h=0.6732$ and $b(z)=1$ for our numerical examples. These results are not sensitive to (scale-independent) bias.

\section{The low \texorpdfstring{$\ell$} ~ spectroscopic \texorpdfstring{$C_\ell(z,\si_z)$'s} ~~: Effects of non-linearities } \label{s:nlin} 
In the previous section we have seen that, even at low $\ell$, spectroscopic $C_\ell(z,\si_z)$'s are affected by the power spectrum at high $k$'s. This implies that they are significantly affected by the non-linearities that exist at high k, even if $k_{\rm Lim}\ll k_{\rm NL}$. Let us introduce the non-linearity scale, as it is often defined, by~\cite{Laureijs:2011gra,Rassat:2008ja}
\be
\sigma(R_{\rm NL},z) = 0.2 \,.
\ee
Here $\sigma^2(R,z)$ is the usual variance of the mass fluctuation in a sphere of radius $R$,
\be
\sigma^2(R,z) \equiv \frac{1}{2\pi^2} \int_0^\infty \frac{dk}{k} \left( \frac{3 j_1(k R)}{k R} \right)^2 k^3P_m(k,z) \,, 
\ee
where $P_m$ is the matter power spectrum.
The amplitude of the matter power spectrum is often parametrized by $\sigma(R=8\mathrm{Mpc}/h,z=0)=\sigma_8$. We associate it with the corresponding non-linearity wave-number,
\be\label{e:kNL}
k_{\rm NL}(z) = \frac{2\pi}{R_{\rm NL}(z)} \, .
\ee

To know the precise change of the angular power spectrum at low $\ell$’s due to non-linearities, one would need to know both the matter and velocity power spectrum precisely at very small (radial) scales and at very large (transverse) scales. In principle, this requires very large, high resolution simulations, including hydrodynamical effects, which goes beyond the scope of the present paper. Here we aim simply to estimate the amount of the modification through answering the following question: By approximately how much are non-linear effects on small scales
 felt at low $\ell$’s in the resulting angular power spectrum, when considering the narrow
window widths associated with spectroscopic surveys?
In order to take into account the uncertainty of modelling, we consider the effects on the spectra of the following three different non-linear models: halofit ~\cite{Takahashi:2012em}, HMcode ~\cite{Mead:2016} and the TNS model ~\cite{Taruya:2010}.

\begin{figure}[ht]
\begin{center}
\includegraphics[width=0.9\linewidth]{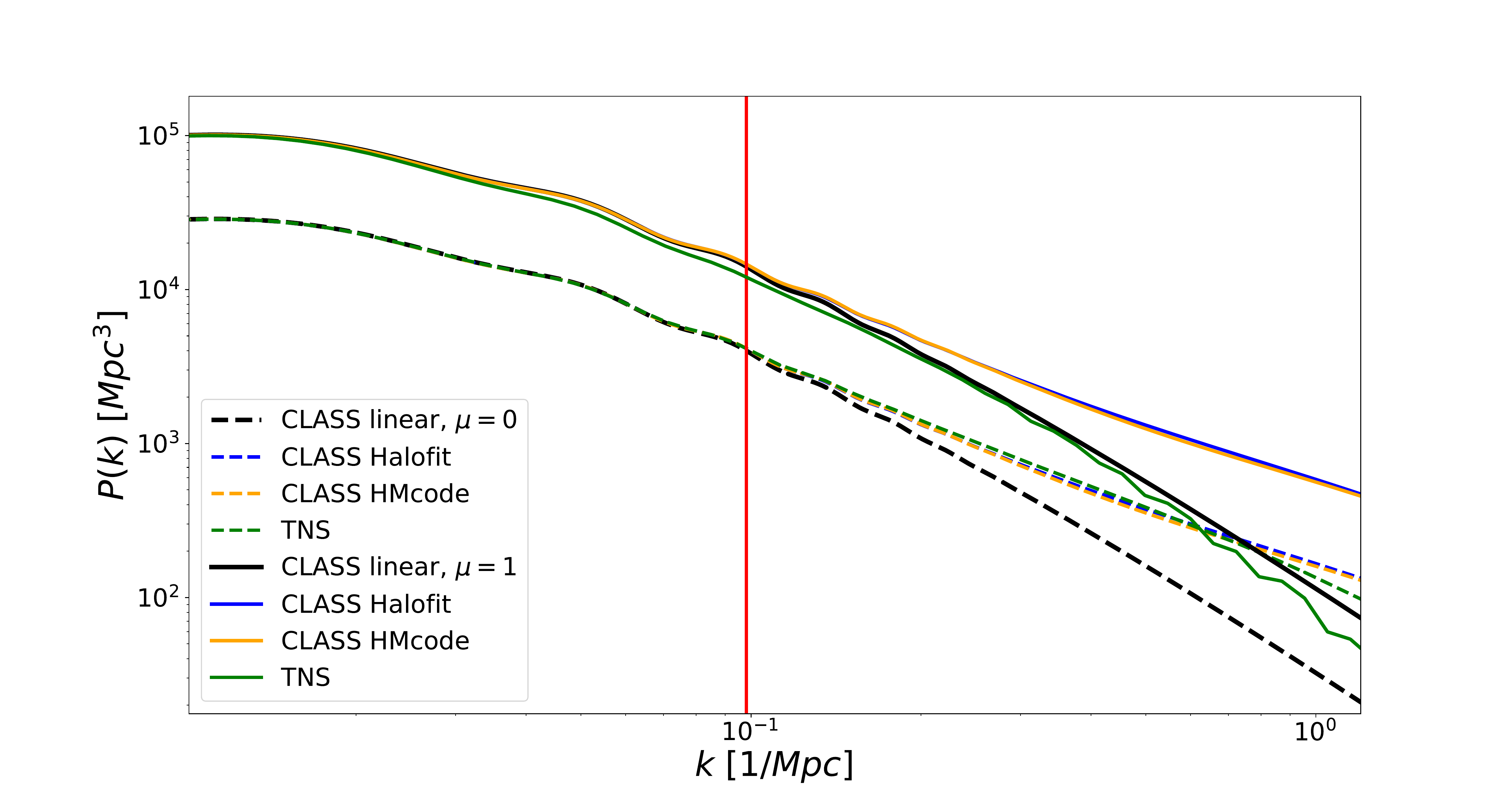}
\end{center}
\caption{\label{f:Pks} We show the standard term contributions 
(density,  rsd and density-rsd cross correlation) of the power spectrum 
$P_{\rm D+RSD}(k,\mu,z) $ 
at $z = 1.0$  with a linear bias b = 1, to examine the effect of various 
non-linear models at small scales. We
plot the spectrum as a function of $k$ for the two values $\mu=0$ and $\mu=1$. The dashed lines indicate
$\mu= 0$ (i.e. density-only contribution), while the solid lines show $\mu=1$. For $\mu = 0$  the non-linear models are very similar, gaining roughly the same amount of power for higher values of k. However, 
for $\mu = 1$, the TNS model differs strikingly, deviating
already at
larger scales, and losing power with respect to the linear spectrum. In red we plot the value of $k_{\rm NL}$ for $z=1$.}
\end{figure}
For a non-linear model of the galaxy power spectrum $P_{gg}$, the velocity divergence power spectrum, $P_{\theta\theta}$ and their correlation, $P_{g\theta}$, the standard density and RSD power spectrum is given by
\be
P_{\rm D+RSD}(k,\mu,z) = P_{gg}+ 2\mu^2/\HH(z)P_{g\theta}(k,\mu,z) + (\mu^2/\HH(z))^2P_{\theta\theta}(k,\mu,z) \,.
\ee
In linear perturbation theory $P_{gg}=b^2(z)P_m$, $P_{\theta\theta}/\HH(z)^2 =f(z)^2P_m$ and $P_{g\theta}/\HH(z) =f(z)b(z)P_m$. For the remainder of this paper, any references to $P(k,\mu,z)$ are taken to mean the linear power spectrum with only standard terms included, i.e. $P_{\rm D+RSD}(k,\mu,z)$, unless otherwise stated.

In Fig.~\ref{f:Pks} we compare the linear power spectrum (black) with various non-linear models, namely: halofit (blue), HMcode (orange) and the TNS model (green). We consider a fixed redshift ($z = 1.0$), and two values of $\mu$: $\mu=0$, for which RSD are absent, is shown as dashed lines, and $\mu=1$, for which RSD are maximal, is shown as solid lines. 
The components of the linear power spectrum correspond, as we have stated, to the standard terms which we consider in this paper. Within linear perturbation theory they are given by \eqref{e:DRSD-lin}. For $\mu = 0$ we are  effectively considering only the contribution from density auto-correlations. In this case all three of the non-linear models are very similar, gaining roughly equal amounts in amplitude. However, for  $\mu=1$ there is a significant difference between the TNS model and the two others. While halofit and HMcode agree so well that their spectra overlie, the TNS model has  much less power on small scales for $\mu=1$, even below the linear spectrum. This comes from the more realistic treatment of the `fingers-of-god' effect in TNS, which reduces significantly the power of the velocity spectrum in the radial direction. This is relatively well modeled in the TNS approach, but neglected in both halofit and HMcode, for which we approximate the density $+$ RSD spectrum by \eqref{e:DRSD-lin}, simply replacing the linear matter power spectrum $P_m$ by the corresponding non-linear model. A comparison between these and several other non-linear approximations and numerical simulations in~\cite{Jalilvand:2019brk} has shown that TNS best models redshift space distortions, while halofit (or HMcode) best models the non-linear matter power spectrum.


\begin{figure}[ht]
\begin{center}
\includegraphics[width=0.9\linewidth]{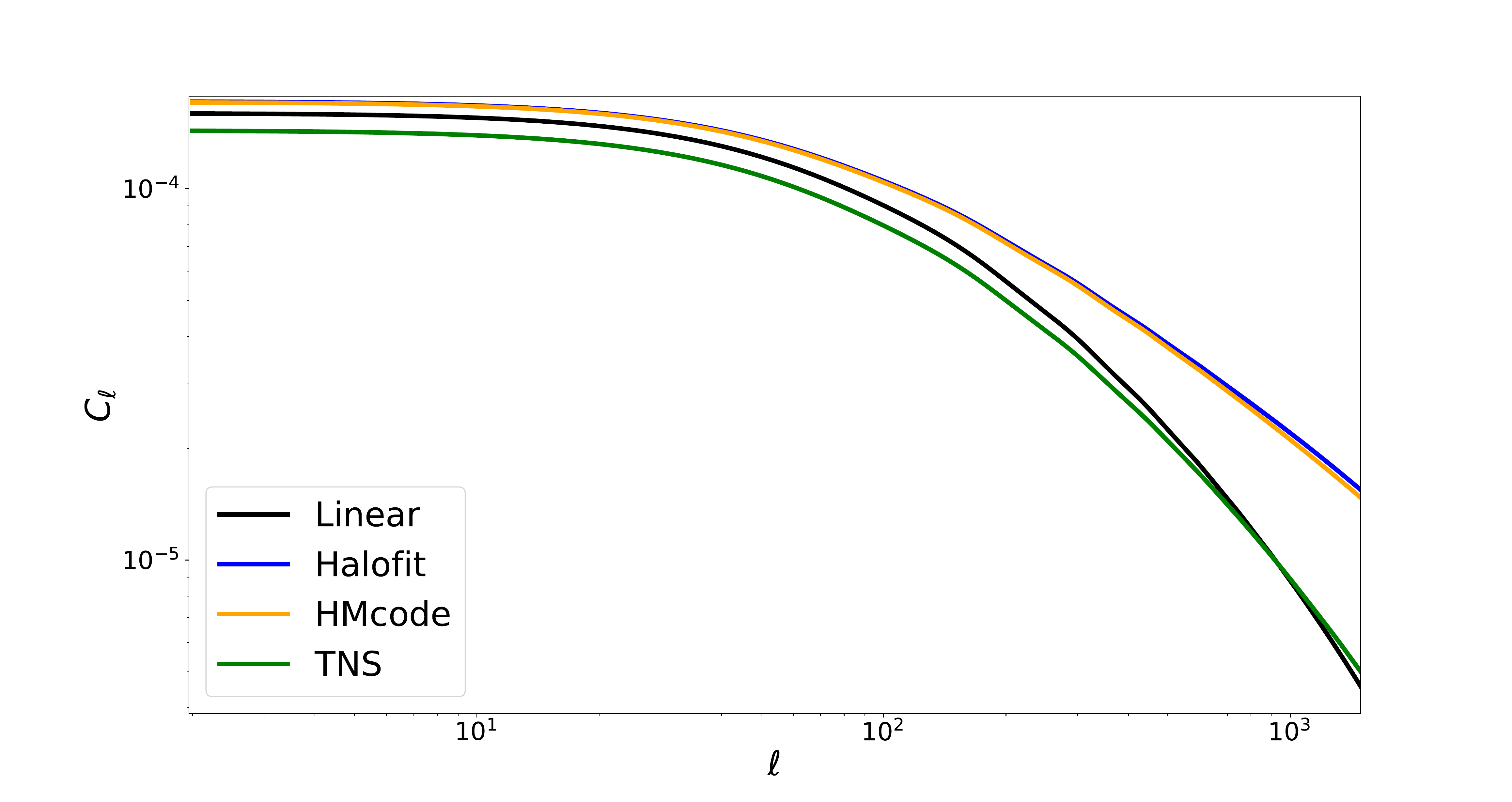}
\end{center}
\caption{\label{f:Clk4a} We show the angular power spectrum $C_\ell(\bar z=1.0,\si_z=10^{-3})$ to illustrate the effect of a narrow window across all scales. We plot the linear power spectrum, and various non-linear spectra, namely those of: Halofit, HMcode and the TNS model}
\end{figure}

In Fig.~\ref{f:Clk4a} we compare the linear $C_\ell$ spectrum (black) with the non-linear halofit (blue), HMcode (orange) and the TNS model (green) spectra for $\bar{z}=1.0$ at fixed spectroscopic bin width $\si_z=10^{-3}$, integrated to a well-converged value of $k_{\max}=5$. The differences in power introduced by the narrow window considered are different depending on the non-linear model chosen, since the effects on the power spectrum $P(k,\mu,z)$ at high $k$ are different. However, in all cases, a constant offset remains even at low $\ell$, indicating that the differences have been propagated across all scales by the narrow window.
\begin{figure}[ht]
\begin{center}
\includegraphics[width=0.9\linewidth]{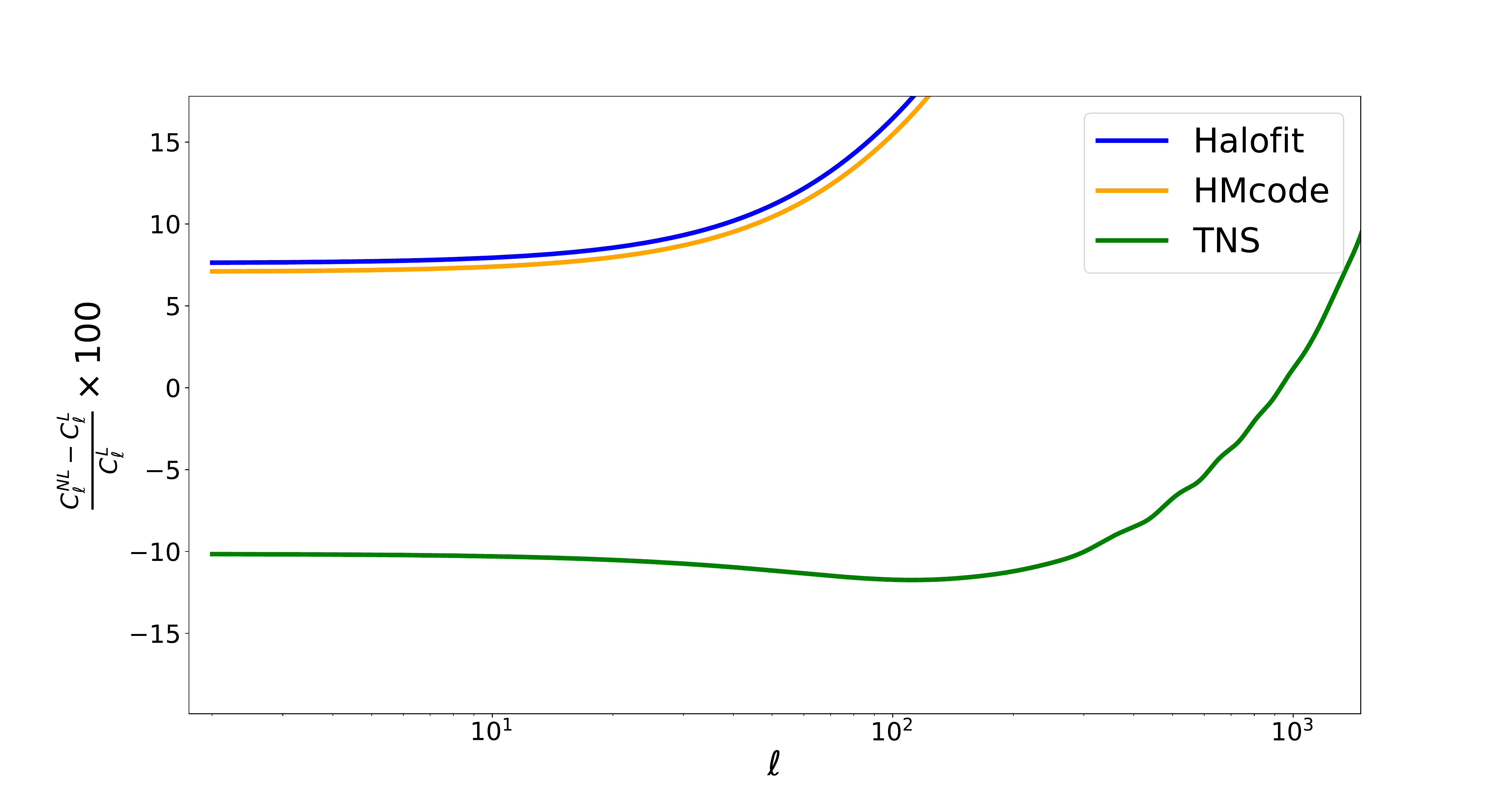}
\end{center}
\caption{\label{f:Clk4b} We show the relative differences between the linear ($C_{\ell}^{\rm L}$) and the non-linear ($C_{\ell}^{\rm NL}$) spectra plotted in Fig.~\ref{f:Clk4a}, in percent. Most importantly, we see that the difference to the linear spectrum at low $\ell$ is a constant offset in the range of $\sim7-10\%$ at $z=1$.}
\end{figure}
This is better visible in  Fig.~\ref{f:Clk4b}, where we show the relative differences, in \%. At $z=1$, spectroscopic power spectra obtained by halofit and HMcode are about $7\%$ larger than the linear perturbation theory result, while the TNS spectrum is about $10\%$ {\em lower} for low $\ell$'s down to $\ell=2$. For halofit and HMcode spectra, at $\ell\gtrsim 100$, where non-linearities are expected to set in, the difference rapidly grows. The relative difference of the TNS spectrum first becomes smaller for $\ell\gtrsim 200$ and overtakes the linear spectrum only at $\ell\simeq 900$, the coincidental point, where the TNS spectrum crosses the linear spectrum. After this point it grows just as rapidly as the other non-linear spectra. In all cases considered, the effects of non-linearities in $P(k,\mu,z)$ (see Fig.~\ref{f:Pks}) at small scales (large $k$), lead to  differences of at least $7\%$ in the $C_\ell$ spectrum at large scales, by virtue of the narrow window investigated. Even though the amplitude and even the sign of
the difference depends on the treatment of non-linearities, we always obtain a constant offset in the $7$ to $10$\% range for $\ell\lesssim 100$ .

For the rest of this analysis, we will use the halofit spectrum as an example to investigate the effect that a boosted amplitude at high $k$ has on the angular power spectrum at low $\ell$. The reason for this, on the one hand, is the simplicity of the model and its good agreement with the more recent HMcode, but also its wide use in the literature. Furthermore, for halofit, non-linearities set in at a somewhat higher value of $k$ than for the TNS model. Therefore, if the angular power spectrum for a low value of $\ell$ is well approximated for a given $k_{\max}$ in the halofit model, this is  also true for the more realistic TNS model.

\begin{figure}[!bh]
\begin{center}\includegraphics[width=0.85\linewidth]{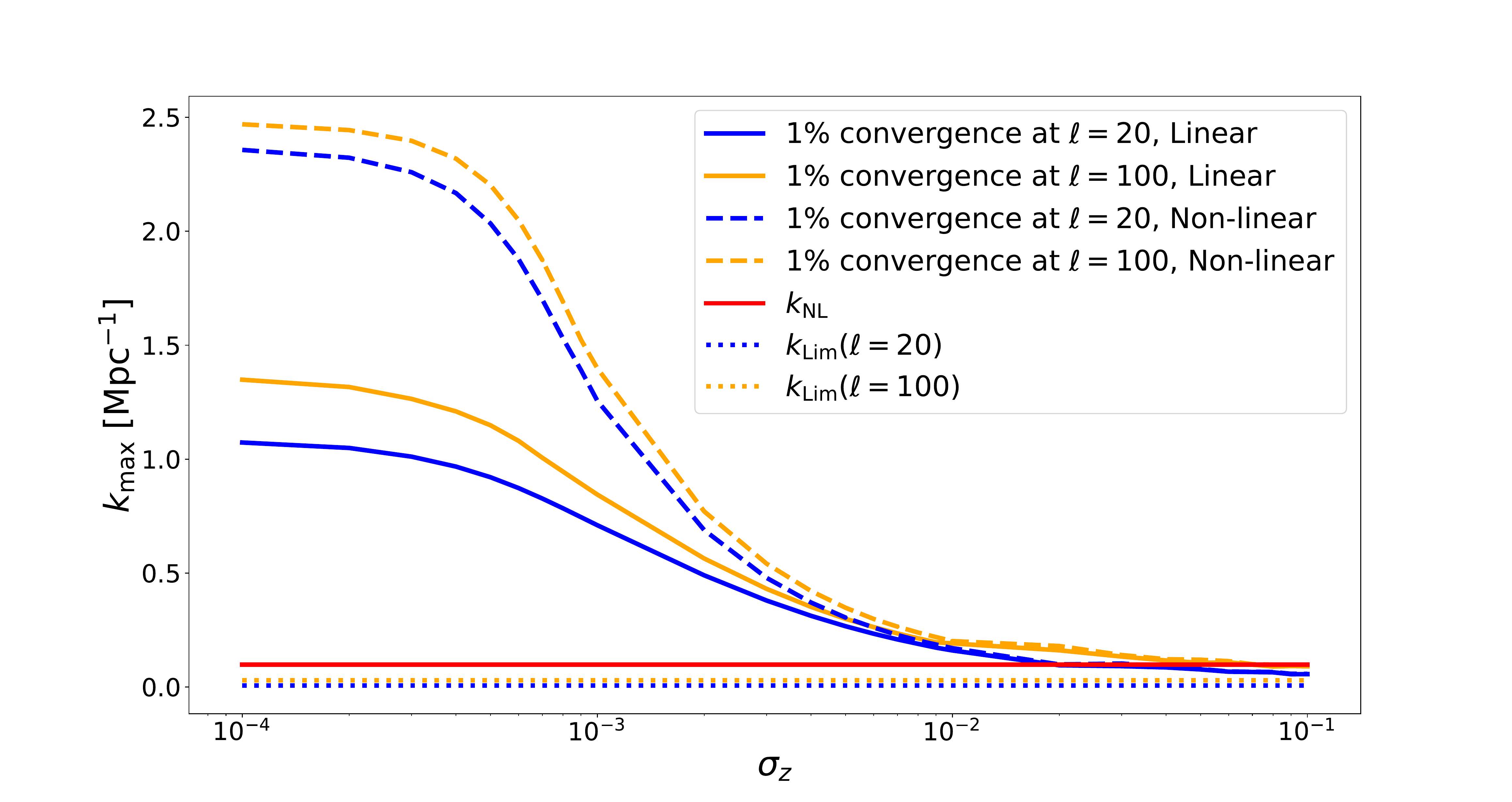}\end{center}
\begin{center}\includegraphics[width=0.85\linewidth]{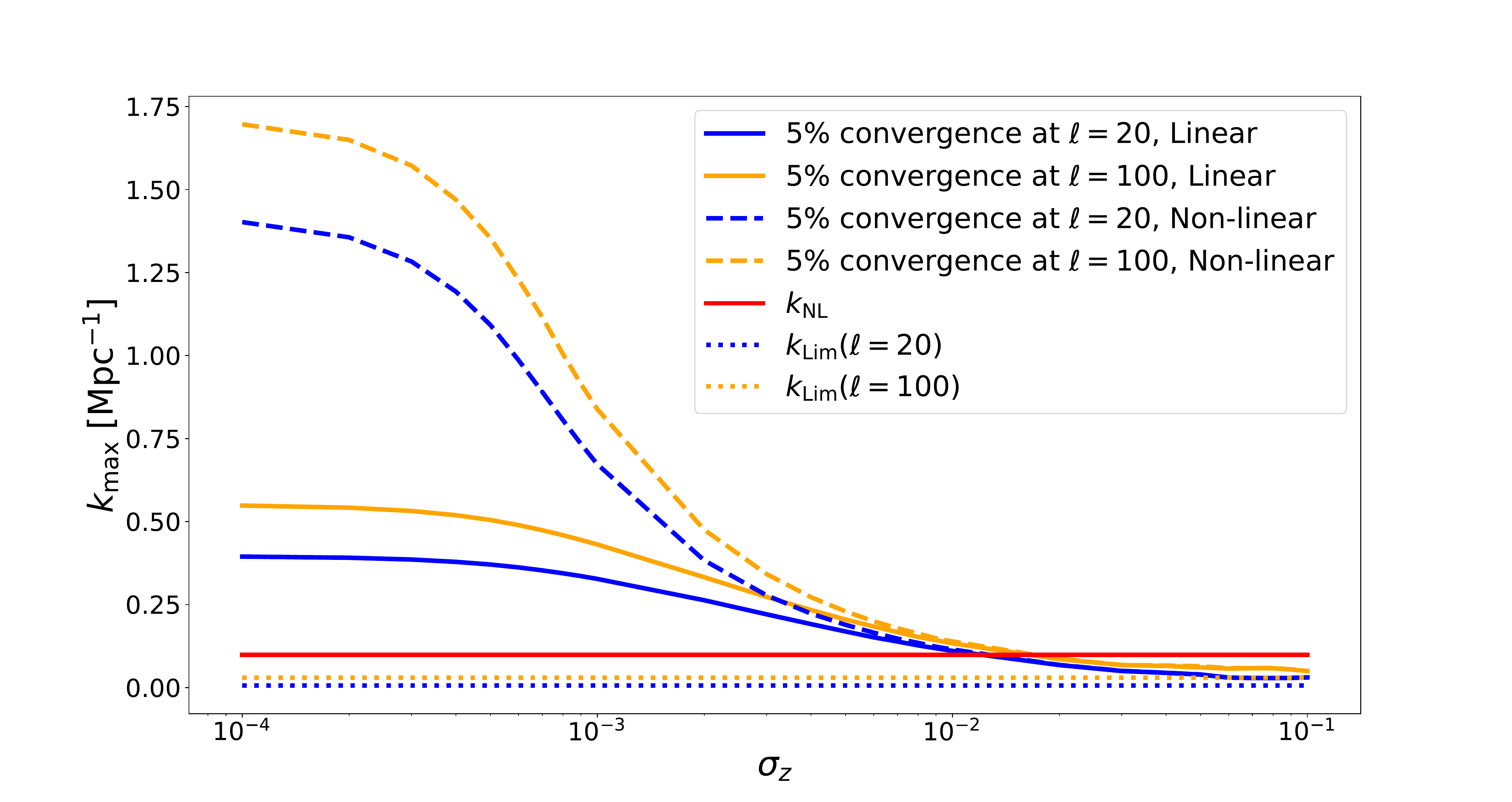}\end{center}
\caption{\label{f:Clk5} We show the values of $k_{\rm max}$ required for $C_\ell(\bar z=1,10^{-4}<\si_z<10^{-1})$ to converge to within $1\%$ (upper panel) or $5\%$ (lower panel) error, relative to the well-converged spectrum calculated with very high $k_{\max}$. We consider the values of the $C_{20}$ (blue) and $C_{100}$ (orange). The solid lines represent the linear spectra, while the dashed lines represent the halofit corrected spectra. We also plot with a red solid line the value of $k_{\rm NL}(z)$ for $z=1$. The dotted lines  indicate the Limber value,  $k_{\rm Lim}(\ell,z)$, with the different colours corresponding to the two respective values of $\ell$ concerned.}
\end{figure}

In Fig.~\ref{f:Clk5} we plot the $k_{\max}$ required to reach a precision of 1\% (upper panel) and 5\% (lower panel) for $C_{20}$ (blue) and 
$C_{100}$ (orange) as a function of $\si_z$. The non-linearity scale is also indicated (red line). At $\si_z=10^{-3}$, $k_{\max}\simeq 10k_{\rm NL}$ for the halofit spectra (dashed), even at $\ell=20$, for 1\% accuracy and somewhat smaller for 5\%.  Up to $\si_z \simeq 10^{-2}$, spectra are affected by more than 5\% by non-linear scales. Since HMcode is very similar to halofit, and for TNS the non-linearities enter at smaller $k$, this is expected to be a safe value for $k_{\max}$ for these models too.  The naive estimate of $k_{\rm max} \sim k_\parallel(z,\sigma_z)$ overestimates the numerically calculated $k_{\rm max}$, especially for smaller windows $\sigma_z$, where the figure shows that the contributions from $k$'s saturate and higher $k$ contributions are not necessary to reach the level of precision considered.

\begin{figure}[!th]
\begin{center}
\includegraphics[width=0.85\linewidth]{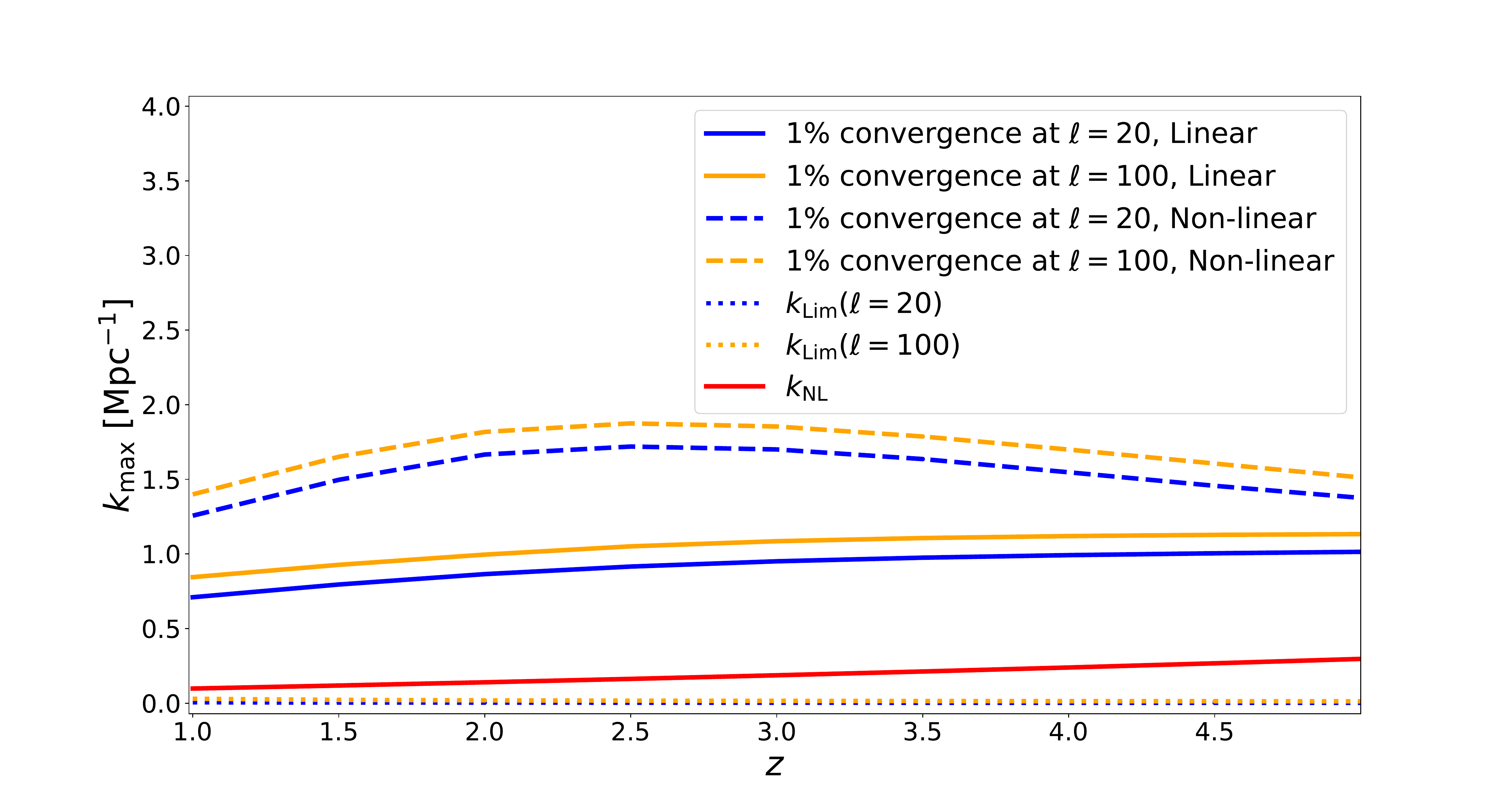}\\
\includegraphics[width=0.85\linewidth]{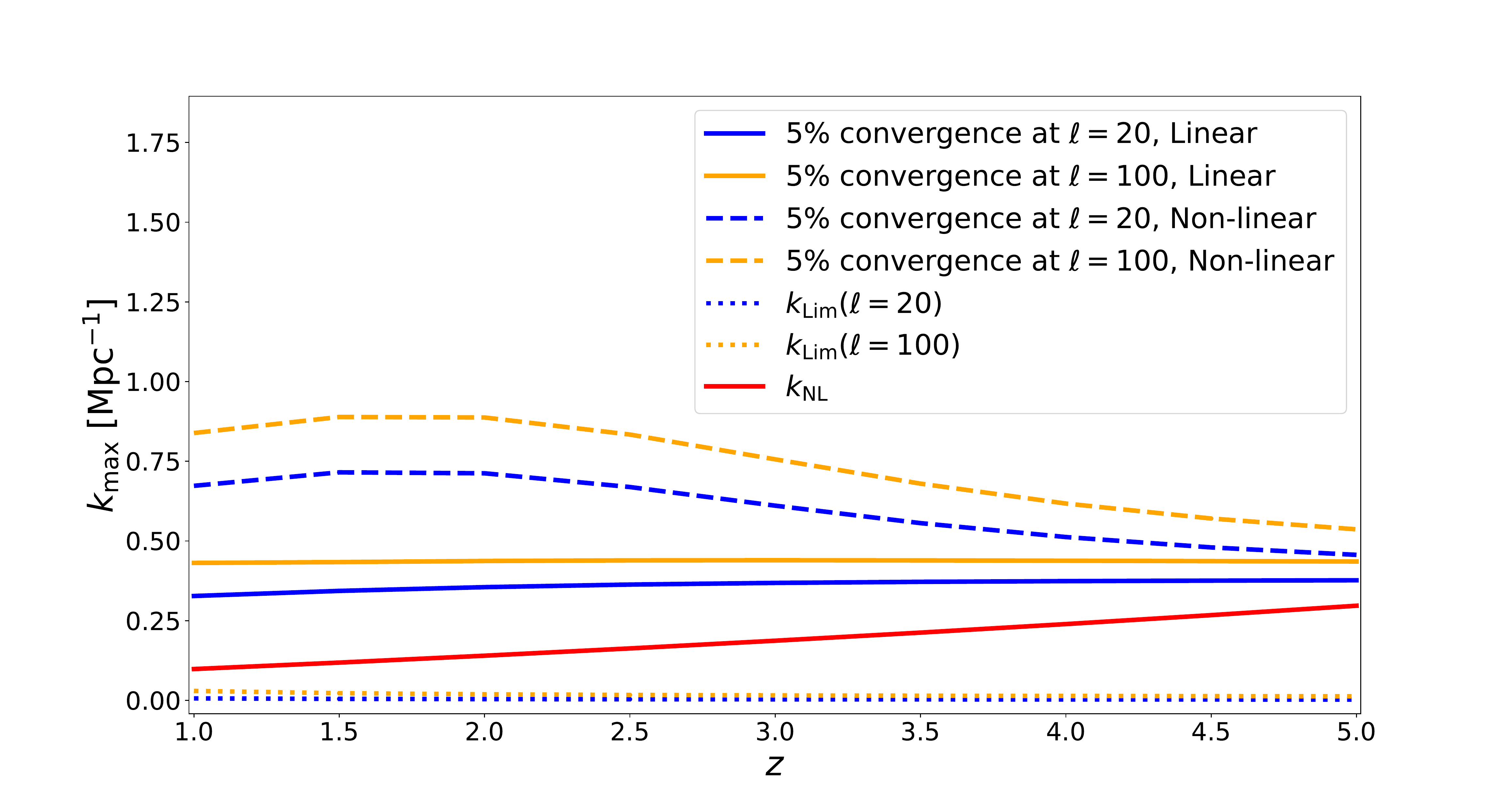}
\end{center}
\caption{\label{f:Clk6} We show the values of $k_{\rm max}$ required for $C_\ell( 1.0<\bar z<5,\si_z=10^{-3})$ to converge to within $1\%$ (upper panel) or $5\%$ (lower panel) error, relative to the spectrum calculated at high k, for $C_{20}$ (blue) and $C_{100}$ (orange). The solid lines represent the linear spectra, while the dashed lines represent the halofit corrections. We plot with a red solid line the value of $k_{\rm NL}(z)$ for the redshift range. The dotted lines indicate the Limber values  $k_{\rm Lim}(z,\ell)$ for the respective values of $\ell$. }
\end{figure}

In Fig.~\ref{f:Clk6} we show the 1\% accuracy values (upper panel) and  5\% accuracy values (lower panel)  of $k_{\max}$ for $\ell=20$ (blue) and $\ell=100$ (orange) for the halofit (dashed) and linear (solid) power spectra at fixed $\si_z=10^{-3}$, typical of a spectroscopic survey, as functions of redshift. We also indicate  $k_{\rm NL}(z)$ (red). The required $k_{\max}(z)$ is larger than $k_{\rm NL}(z)$ throughout. Non-linearities still affect the spectra up to $z=5$ by more  than 5\%. Note also that  $k_{\max}$ depends only weakly on redshift for the linear power spectrum. Our naive estimates $k_{\max}\sim k_{\pa}(z,\si_z)=2\pi H(z)/\si_z \simeq 2.5(H(z)/H(1))/{\rm Mpc}$  (which {again} overestimates the convergence scale in all cases) would be rising with redshift. The near constancy of $k_{\max}$ is caused by the fact that the matter power spectrum decreases for $k>k_{\rm eq}\sim 0.01/$Mpc. The rise in $k_{\pa}(z,\si_z)$ and the decrease in the matter power spectrum  seem to nearly balance each other in the linear power spectrum. However, the redshift dependence of  $k_{\max}$ becomes quite pronounced for the non-linear (halofit) spectrum. It is therefore purely a consequence of the shape of the power spectrum, which decreases much less steeply when considering the non-linear case. The value of $k_{\max}(z)$ for the halofit power spectrum first rises until about $z=2$, (a result of the competing effects, with increasing redshift, of weakening non-linearities and the decreasing scale of modes probed by a fixed window width), and then decreases towards higher redshifts, slowly approaching the value for the linear power spectrum, since non-linearities become weaker.

\section{Cross-correlations, \texorpdfstring{$C_\ell(z_1,z_2)$ for $z_1\neq z_2$} ~ }\label{s:cross} 

In this section we consider cross-correlations of the linear $C_\ell(z_1,z_2)$, with $z_1\neq z_2$. It is well known that for widely-separated redshifts, cross-correlations are dominated by the lensing--density (at low redshifts) or the lensing--lensing (at high redshift) terms~\cite{Montanari:2015rga}. Furthermore, these terms are well approximated with the Limber approximation, to which only the spectrum at $k_{\rm Lim}=(\ell+1/2)/r(z)$ contributes~\cite{Matthewson:2020rdt}.
Here we study just the standard terms, since only these may be affected by small scale contributions.
The relation between the standard terms and the lensing terms depends on both galaxy bias $b(z)$ and magnification bias $s(z)$, which are survey dependent, see e.g.~\cite{DiDio:2013bqa}. Therefore, we do not discuss their relative amplitude in this general study. We consider only redshift pairs $(z_1,z_2)$ which are not too widely separated, so as to avoid cases where the contribution of the standard terms is completely negligible.
In the Limber approximation, which is well known to fail for the standard terms in narrow redshift bins~\cite{DiDio:2018unb,Matthewson:2020rdt}, these terms simply vanish if the bins are not overlapping.

\begin{figure}[ht]
\begin{center}
\includegraphics[width=0.85\linewidth]{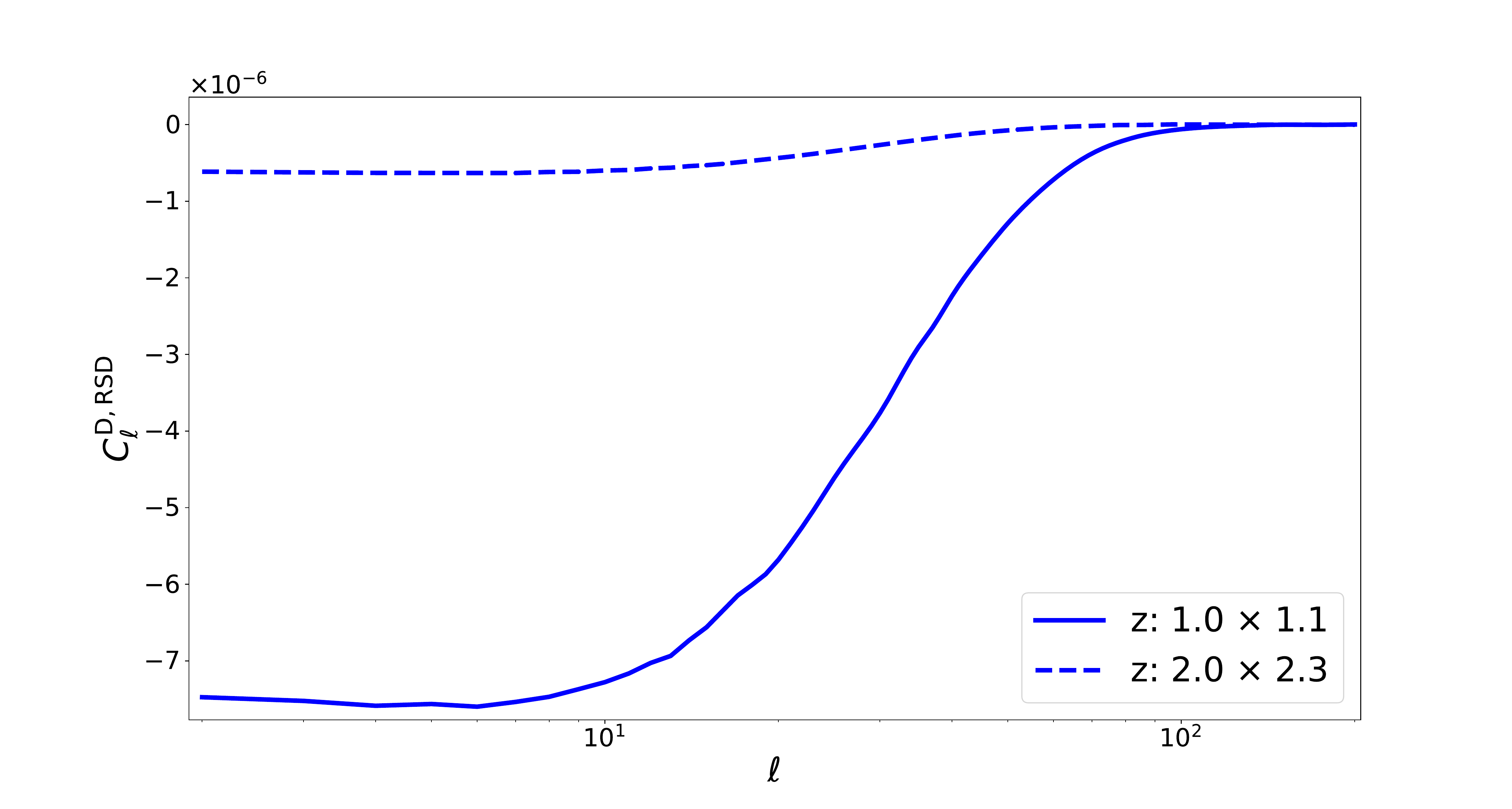}
\end{center}
\caption{\label{f:X1-0.1spec} We show the angular power spectrum for the redshift pairs: ($z_1=1$, $z_2=1.1$) and ($z_1=2$, $z_2=2.3$). Only density and redshift space distortions are included in the signal here, and a window of $\sigma_z = 10^{-3}$ is chosen. At $\ell \gtrsim 100$ the contribution from these terms is much less significant.} 
\end{figure}

To obtain some intuition\footnote{This approximation is not very accurate if $z_1$ and $z_2$ are not very close and we shall not use it for our numerical results in this section, using CAMB sources instead.}, we consider again the flat sky approximation and integrate over tophat windows of width $\si_z$ centered at $z_1$ and $z_2$. Setting $\bar z= (z_1+z_2)/2$, a short calculation gives
\be\label{e:ClWx}
C_\ell(z_1,z_2,\si_z)\simeq \frac{1}{2\pi r^2(\bar z)}\int_{-\infty}^\infty dk_\pa P(k,\bar z)\exp\left(ik_\pa\left(\frac{z_1}{H(z_1)}-\frac{z_2}{H(z_2)}\right)\right)j_0^2\left(\frac{k_\pa\si_z}{2H(\bar z)}\right) \,.
\ee
In this approximation we have used that $H(z)$ and $P(k,z)$ change slowly with redshift, while the exponential changes rapidly. In addition to the low-pass filter $k_\pa(z,\si_z)=2\pi H(\bar z)/\si_z$, we now have a new scale given by the wave number $k_\times(z_1,z_2)$, after which the exponential has performed about one oscillation,
\be
k_\times(z_1,z_2)= \frac{2\pi H(\bar z)}{|z_2-z_1|} < \frac{2\pi H(\bar z)}{\si_z}= k_\pa(z,\si_z) \,.
\ee
We assume that the two redshift bins are not overlapping, i.e. $|z_2-z_2|> \si_z$. As a first guess one might hope that contributions from values of $k_\pa>k_\times$ are averaged out by oscillations and can be neglected. The situation is actually quite interesting. To analyze it, let us first assume $P(k,z)$ to be independent of $k_\pa$, a reasonable approximation for large $\ell$, where $k\simeq \ell/r(z)$. In this case we can integrate \eqref{e:ClWx} exactly. Defining
\be
F(a) =\int_{-\infty}^\infty \hspace{-0.2cm}dx\, e^{ix}j_0^2(ax) \;=\; \left\{
\begin{array}{ll} 0\,~ & a\leq 1/2 \\ \frac{\pi}{2a^2}(2a-1)\, ~ & a\geq 1/2\,,
\end{array} \right.
\ee
we  obtain
\be
C_\ell(z_1,z_2,\si_z)\simeq \frac{H(\bar z)}{2\pi|z_1-z_2| r^2(\bar z)}P(\ell/r(\bar z),\bar z) F\left(\frac{\si_z/2}{|z_1-z_2|}\right)\,,
\ee
which vanishes entirely for well separated windows, $\si_z<|z_1-z_2|$. This corresponds to our findings for large $\ell>100$, see Fig.~\ref{f:X1-0.1spec}, where a window size $\si_z=10^{-3}$ is chosen. For smaller $\ell$ the situation is more complicated and the result depends entirely on the behavior of $P(k,z)$.
Our numerical examples show that, depending on the redshift pair considered, the value $k_{\max}$ required to achieve an accuracy of 10\% for the cross-spectrum of the standard terms can be much smaller than $k_\times$ or
 up to more than 8 times higher.

\begin{figure}[ht]
\includegraphics[width=0.85\linewidth]{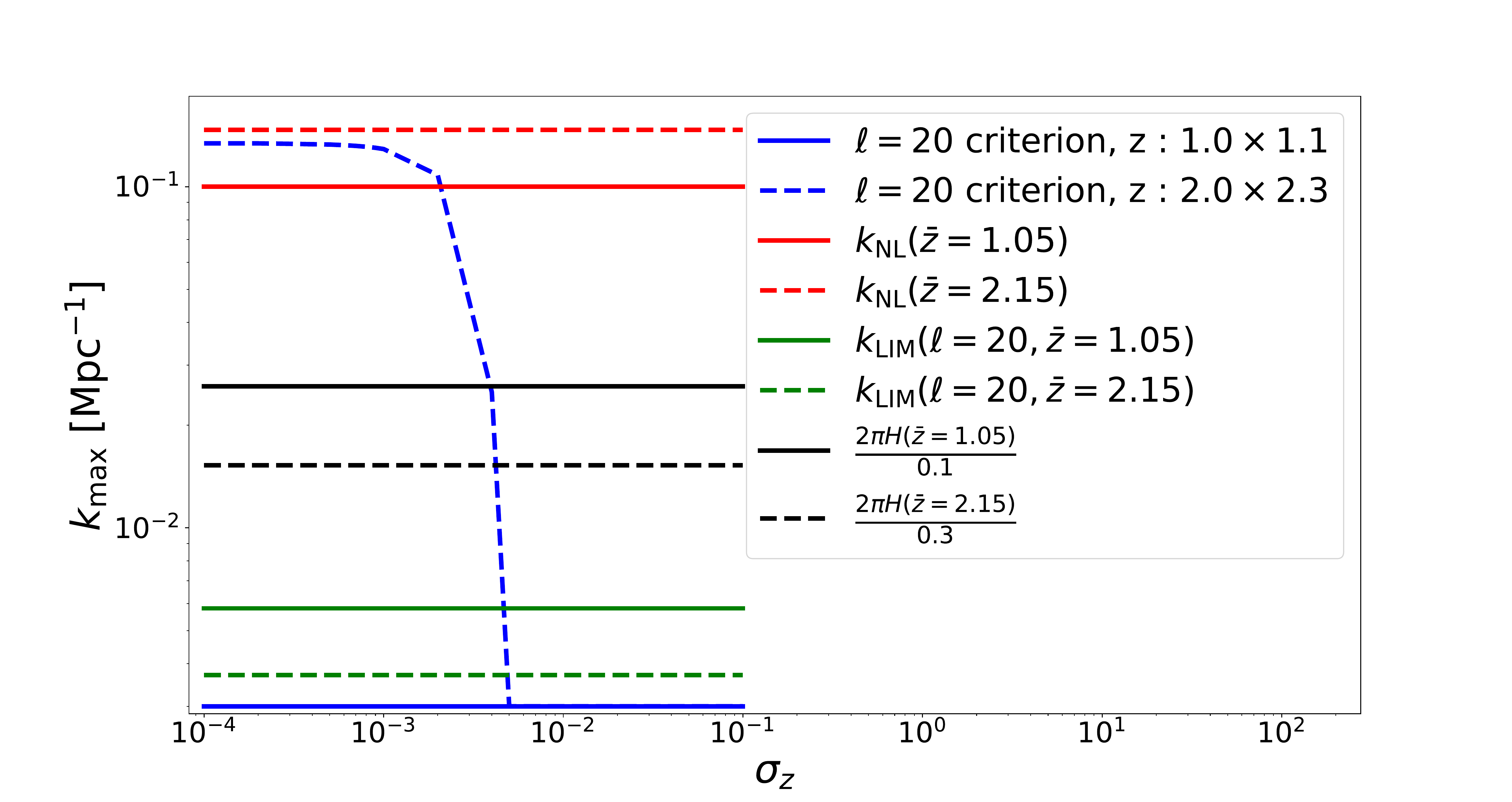}
\caption{\label{f:X1-0.1} We show the value of $k_{\max}$ needed to achieve a 10\% precision for the cross-correlation spectrum $C_{20}(z_1,z_2)$ at ($z_1=1$, $z_2=1.1$) and ($z_1=2$, $z_2=2.3$) as a function of the window width $\si_z$. Here only density and redshift space distortions are included in the signal.} 
\end{figure}

The contribution from the standard terms in the cross-correlations are very small and tend to zero for $\ell\gtrsim 100$, see Fig.~\ref{f:X1-0.1spec}.  Thus, measuring these terms with observations will be quite challenging. For this reason, we only investigate the $k_{\max}$ needed for a 10\% accuracy in this case. Also, as the signal is very close to zero at $\ell=100$, we concentrate on the case $\ell=20$ which is also more relevant here, as it is typically expected to converge for smaller values of $k_{\max}$.

The $k_{\max}$ needed to achieve 10\% accuracy in the linear cross-correlation spectrum is about $8k_\times$ for $(z_1,z_2)=(2,2.3)$ and $\ell=20$; however, it remains smaller than $k_{\rm NL}$. Thus we deduce that, even though a narrower window function still increases the $k_{\max}$ required for good convergence, it is not very significant and  non-linearities are much less important at low $\ell$ in the case of cross-correlations, see Fig.~\ref{f:X1-0.1}. For $(z_1,z_2)=(1,1.1)$, $k_{\max}$  is much smaller than $k_\times$, actually roughly given by $k_{\rm Lim}$, which would be in agreement with our naive expectation. It seems then that the scale $k_\times$ does not characterise the scale of important contributions well, if at all. While the reason for this is not entirely clear to the authors, one point is simply that the cross-correlation signal of density and redshift space distortion for $(z_1,z_2)=(2,2.3)$ is about 10 times smaller than the one for  $(z_1,z_2)=(1,1.1)$ for $\ell=20$, making an accuracy of $10\%$ even more difficult to attain. 
We also found that, when considering higher redshifts, e.g. $\bar z\sim 3$, it is possible for $k_{\max}$ to decrease again to less than $k_\times$ and, when considering a smaller separation $(z_1,z_2)=(2,2.2)$, where the standard terms are larger, we find a very small $k_{\max}$, once more in agreement with our naive expectation.

However, the accuracy of the standard terms is in any case not so relevant for cross-correlation spectra, something which should be borne in mind when interpreting the values of $k_{\rm max}$ in this section.  First of all, as already mentioned, the lensing signal cannot be neglected in these spectra and including it significantly enhances the total signal, {increasing the accuracy in calculating the total signal of the Limber approximation, which requires a lower $k_{\max}$.}
\begin{figure}[!ht]
\begin{center}
\includegraphics[width=0.85\linewidth]{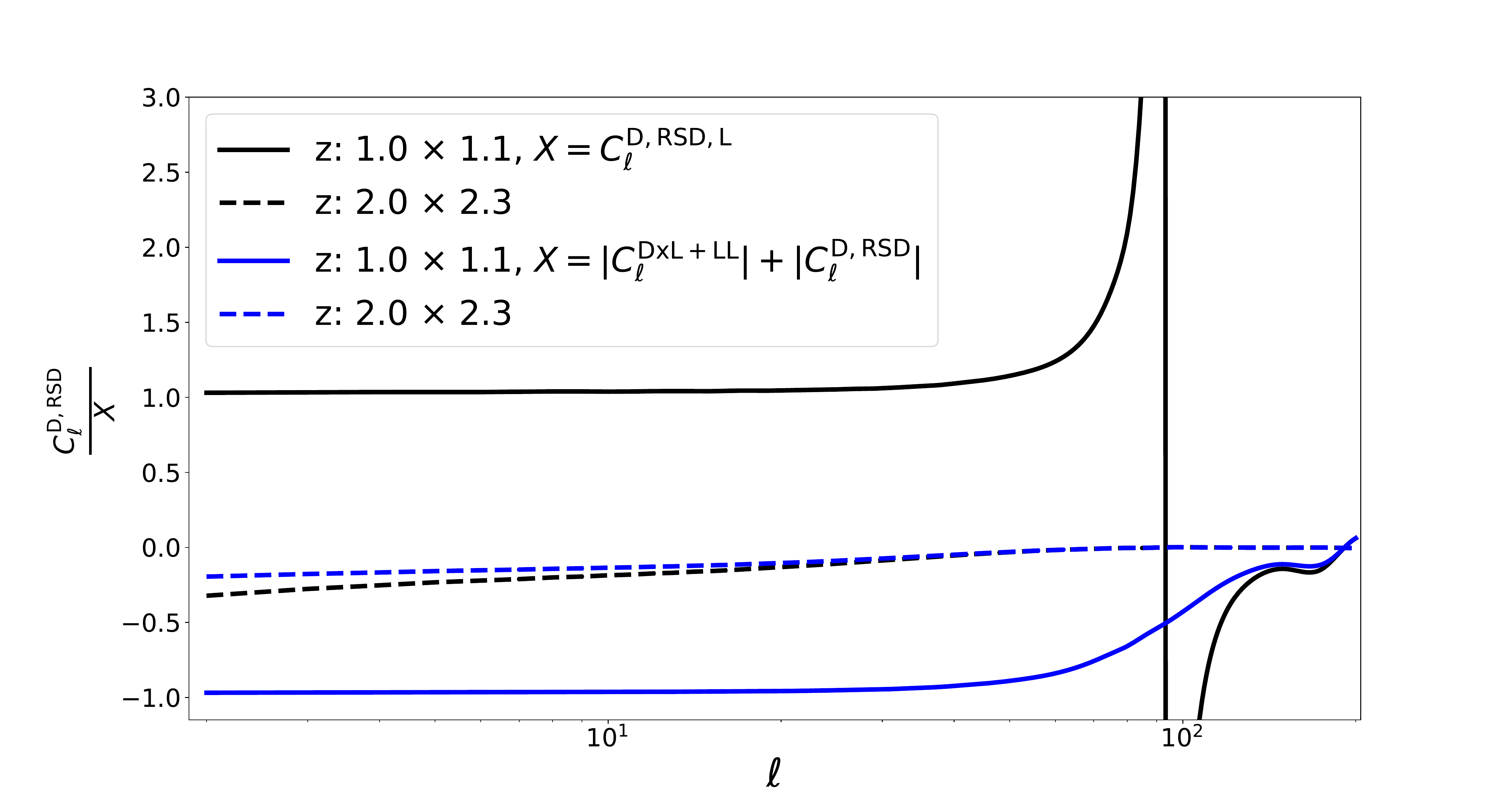}
\end{center}
\caption{\label{f:SKA} We show the fraction of the full angular power spectrum made up for by the density and redshift space distortion terms, in the case of an SKAII-like spectroscopic galaxy survey (black). Since the total signal vanishes at $\ell\sim 100$, we also show $C_\ell^{D,RSD}/( |C_\ell^{\rm D,RSD}|+|C_\ell^{\rm L\times D + LL}|)$ (blue).} 
\end{figure}
In Fig.~\ref{f:SKA} we show as an example the ratio of the standard terms, $C_\ell^{\rm D,RSD}$, to the total spectrum including also the lensing terms, using values for $b(z)$ and $s(z)$ for SKA{II} as given in~\cite{Jelic-Cizmek:2020pkh}, Appendix A4. We use the same redshift pairs as for the previous figures (black lines). For $(z_1,z_2)=(1,1.1)$, at low $\ell$, the ratio is slightly larger than 1, since the lensing contribution, dominated by the density-lensing cross-correlation, is negative. The divergence at $\ell\simeq 100$ is the consequence of a zero-crossing of the denominator, the total spectrum. To remove it we  also plot  $C_\ell^{\rm D,RSD}/(|C_\ell^{\rm D,RSD}|+|C_\ell^{\rm L\times D+LL}|)$ (blue lines). At  $\ell>120$, the total signal is positive, the lensing term now dominates and the standard terms contribute less than about 20\%. For  $(z_1,z_2)=(2,2.3)$ the lensing term is always positive and dominates. The standard terms contribute only about 13\% already for $\ell\gtrsim 20$, so that a 10\% error in the standard terms leads to a  1.3\% error in 
the total result. Assuming a numerical accuracy of about 1\% for these cross-correlations, it is not clear whether the high value of $k_{\max}$ found in this case should not be (at least partially) attributed to numerical error in the CAMB-code with which the spectra have been computed. For $(z_1,z_2)=(2,2.2)$ the standard terms still contribute nearly 50\% at $\ell=20$ and the required $k_{\max}$ is again very small, while for $(z_1,z_2)=(2,2.5)$  they are below 5\%{, while the required $k_{\rm max}$ is even larger than that for $(z_1,z_2) = (2,2.3)$}. But as mentioned above, these percentages depend on $b(z)$ and $s(z)$.

The second reason why the accuracy of the standard terms is not of utmost importance, is that cross-correlations cannot be measured as accurately as auto-correlations, due to cosmic variance. The cosmic variance of cross-correlations is dominated by the much larger auto-correlations, 
\be
{\rm Var}\left(C_\ell(z_1,z_2)\right) \simeq \frac{C_\ell(z_1,z_1)C_\ell(z_2,z_2)}{2\ell+1} \,.
\ee
 For example at $z\simeq 1$, $\De z \geq 0.1$ and $\ell\leq 50$ auto-correlations are of the order of $10^{-4}$ while cross-correlations are about $10^{-7}$. Hence, for a given redshift combination and $\ell\sim 20$, we expect a signal-to-noise ratio from the standard term alone of less than $0.1$. To overcome this large cosmic variance, one will have to consider significant binning in $\ell$- and $z$-space.

\section{Conclusion}\label{s:con}
In this paper we have studied the angular power spectra for galaxy number counts, $C_\ell(z,z')$. In particular, we have investigated their dependence on the widths of the considered redshift window, $\si_z$. We have found that in slim windows, relevant for spectroscopic surveys, i.e. $\si_z\simeq 10^{-3}$, the spectra are strongly  affected by high values of $k$, even for very low multipoles. 
This means that for spectroscopic redshift bins, non-linearities affect the angular power spectra  all the way down to $\ell=2$. More precisely, at low  $\ell$ they produce a constant offset in the 10\% range. This is especially true in the case of equal redshifts, $z=z'$, and much less so for cross-correlations of different redshifts that are non-overlapping, but still close enough so that the standard density+RSD terms in the number counts are nevertheless considerable. Even though the exact effect of non-linearities is not really established with this work, which does not perform N-body simulations, differences of up to 17\% are observed, depending on the non-linear model (e.g. between HMcode and TNG). Spectroscopic surveys are very important to measure the growth rate of perturbations, one of the most interesting variables to discriminate between different dark energy models~\cite{Cooray:2003hd,DiPorto:2007ovd,Jelic-Cizmek:2020pkh}. 

Our finding is an additional motivation to use, instead of the angular power spectrum, the angular correlation function for larger redshift bins and the angular power spectrum only for cross-correlations of these bins. The angular correlation function is not affected by non-linearities for sufficiently large separations, in the analysis of spectroscopic surveys. This has been advocated already in the past~\cite{Tansella:2017rpi,Tansella:2018sld}. In Ref.~\cite{Tansella:2018sld} a public code for the fully-relativistic correlation function is presented and a significant speedup of this code has been achieved recently~\cite{jeliccizmek:2020fl}. Another argument for why the angular power spectra are not suited to spectroscopic redshift resolution is the fact that even in very big surveys we would at best have a few thousand galaxies per redshift bin with $\si_z\simeq 10^{-3}$ which leads to very substantial shot noise.

For cross-correlations, the value of $k_{\max}$ needed to achieve a precision of 10\% for the density and RSD contribution to the $C_\ell(z,z')$ is not so high due to the following reasons: 1) 
The cross-correlation signal has an additional oscillating function with a phase proportional to the redshift difference in its integral, which reduces the contributions from high $k_\pa$.  2)
The lensing terms of cross-correlations, that are well approximated by the Limber approximation, are much more significant than their contribution to  auto-correlations and so the $k_{\max}$ needed for a 10\% precision of the total, measured power spectra is significantly smaller. 3) Finally, the cross-correlation signal is much smaller than the auto-correlation signal, while its cosmic variance is similar. Therefore, we expect to achieve significantly lower precision in the measurements of cross-correlations. For these reasons, we have determined the value $k_{\max}$ required to achieve 10\% precision for the contribution of the standard terms to cross-correlations.
At large redshift differences, cross-correlations are dominated by the lensing term and one can safely apply the Limber approximation to compute them, see~\cite{Matthewson:2020rdt} for a detailed study. 

We therefore consider our results very important for the auto-correlations which will be measured with a high signal-to-noise ratio in upcoming spectroscopic surveys like Euclid or SKA, but not so relevant for cross-correlations of non-overlapping redshift bins.

\section*{Acknowledgements}
{It is a pleasure to thank Martin Kunz and Francesca Lepori for enlightening discussions.  The authors acknowledge financial support from the Swiss National Science Foundation grant no. 200020-182044}

 \bibliographystyle{JHEP}
\bibliography{refs}

\providecommand{\href}[2]{#2}\begingroup\raggedright\begin{thebibliography}{10}

\bibitem{Bonvin:2011bg}
C.~Bonvin and R.~Durrer, {\it {What galaxy surveys really measure}},  {\em
  Phys. Rev.} {\bf D84} (2011) 063505,
  [\href{http://arxiv.org/abs/1105.5280}{{\tt arXiv:1105.5280}}].

\bibitem{Challinor:2011bk}
A.~Challinor and A.~Lewis, {\it {The linear power spectrum of observed source
  number counts}},  {\em Phys. Rev.} {\bf D84} (2011) 043516,
  [\href{http://arxiv.org/abs/1105.5292}{{\tt arXiv:1105.5292}}].

\bibitem{Jeong:2011as}
D.~Jeong, F.~Schmidt, and C.~M. Hirata, {\it {Large-scale clustering of
  galaxies in general relativity}},  {\em Phys. Rev.} {\bf D85} (2012) 023504,
  [\href{http://arxiv.org/abs/1107.5427}{{\tt arXiv:1107.5427}}].

\bibitem{DiDio:2013bqa}
E.~Di~Dio, F.~Montanari, J.~Lesgourgues, and R.~Durrer, {\it {The CLASSgal code
  for Relativistic Cosmological Large Scale Structure}},  {\em JCAP} {\bf 1311}
  (2013) 044, [\href{http://arxiv.org/abs/1307.1459}{{\tt arXiv:1307.1459}}].

\bibitem{DiDio:2013sea}
E.~Di~Dio, F.~Montanari, R.~Durrer, and J.~Lesgourgues, {\it {Cosmological
  Parameter Estimation with Large Scale Structure Observations}},  {\em JCAP}
  {\bf 1401} (2014) 042, [\href{http://arxiv.org/abs/1308.6186}{{\tt
  arXiv:1308.6186}}].

\bibitem{Alonso:2015uua}
D.~Alonso, P.~Bull, P.~G. Ferreira, R.~Maartens, and M.~Santos, {\it {Ultra
  large-scale cosmology in next-generation experiments with single tracers}},
  {\em Astrophys. J.} {\bf 814} (2015), no.~2 145,
  [\href{http://arxiv.org/abs/1505.07596}{{\tt arXiv:1505.07596}}].

\bibitem{Lorenz:2017iez}
C.~S. Lorenz, D.~Alonso, and P.~G. Ferreira, {\it {The impact of relativistic
  effects on cosmological parameter estimation}},
  \href{http://arxiv.org/abs/1710.02477}{{\tt arXiv:1710.02477}}.

\bibitem{Jalilvand:2019brk}
M.~Jalilvand, B.~Ghosh, E.~Majerotto, B.~Bose, R.~Durrer, and M.~Kunz, {\it
  {Nonlinear contributions to angular power spectra}},  {\em Phys. Rev. D} {\bf
  101} (2020), no.~4 043530, [\href{http://arxiv.org/abs/1907.13109}{{\tt
  arXiv:1907.13109}}].

\bibitem{Limber:1954zz}
D.~N. Limber, {\it {The Analysis of Counts of the Extragalactic Nebulae in
  Terms of a Fluctuating Density Field. II}},  {\em Astrophys. J.} {\bf 119}
  (1954) 655.

\bibitem{LoVerde:2008re}
M.~LoVerde and N.~Afshordi, {\it {Extended Limber Approximation}},  {\em Phys.
  Rev.} {\bf D78} (2008) 123506, [\href{http://arxiv.org/abs/0809.5112}{{\tt
  arXiv:0809.5112}}].

\bibitem{Assassi:2017lea}
V.~Assassi, M.~Simonovi{\'c}, and M.~Zaldarriaga, {\it {Efficient Evaluation of
  Cosmological Angular Statistics}},  {\em JCAP} {\bf 1711} (2017), no.~11 054,
  [\href{http://arxiv.org/abs/1705.05022}{{\tt arXiv:1705.05022}}].

\bibitem{Matthewson:2020rdt}
W.~L. Matthewson and R.~Durrer, {\it {The Flat Sky Approximation to Galaxy
  Number Counts}},  {\em JCAP} {\bf 02} (2021) 027,
  [\href{http://arxiv.org/abs/2006.13525}{{\tt arXiv:2006.13525}}].

\bibitem{Durrer:2020fza}
R.~Durrer, {\em {The Cosmic Microwave Background}}.
\newblock Cambridge University Press, 2nd~ed., 2020.

\bibitem{Blas:2011rf}
D.~Blas, J.~Lesgourgues, and T.~Tram, {\it {The Cosmic Linear Anisotropy
  Solving System (CLASS) II: Approximation schemes}},  {\em JCAP} {\bf 1107}
  (2011) 034, [\href{http://arxiv.org/abs/1104.2933}{{\tt arXiv:1104.2933}}].

\bibitem{Takahashi:2012em}
R.~Takahashi, M.~Sato, T.~Nishimichi, A.~Taruya, and M.~Oguri, {\it {Revising
  the Halofit Model for the Nonlinear Matter Power Spectrum}},  {\em Astrophys.
  J.} {\bf 761} (2012) 152, [\href{http://arxiv.org/abs/1208.2701}{{\tt
  arXiv:1208.2701}}].

\bibitem{Laureijs:2011gra}
{\bf Euclid} Collaboration, R.~Laureijs et~al., {\it {Euclid Definition Study
  Report}},  \href{http://arxiv.org/abs/1110.3193}{{\tt arXiv:1110.3193}}.

\bibitem{Rassat:2008ja}
A.~Rassat, A.~Amara, L.~Amendola, F.~J. Castander, T.~Kitching, M.~Kunz,
  A.~Refregier, Y.~Wang, and J.~Weller, {\it {Deconstructing Baryon Acoustic
  Oscillations: A Comparison of Methods}},
  \href{http://arxiv.org/abs/0810.0003}{{\tt arXiv:0810.0003}}.

\bibitem{Mead:2016}
A.~J. Mead, C.~Heymans, L.~Lombriser, J.~A. Peacock, O.~I. Steele, and H.~A.
  Winther, {\it Accurate halo-model matter power spectra with dark energy,
  massive neutrinos and modified gravitational forces},  {\em Monthly Notices
  of the Royal Astronomical Society} {\bf 459} (Mar, 2016) 1468–1488.

\bibitem{Taruya:2010}
A.~Taruya, T.~Nishimichi, and S.~Saito, {\it Baryon acoustic oscillations in
  2d: Modeling redshift-space power spectrum from perturbation theory},  {\em
  Physical Review D} {\bf 82} (Sep, 2010).

\bibitem{Montanari:2015rga}
F.~Montanari and R.~Durrer, {\it {Measuring the lensing potential with
  tomographic galaxy number counts}},  {\em JCAP} {\bf 1510} (2015), no.~10
  070, [\href{http://arxiv.org/abs/1506.01369}{{\tt arXiv:1506.01369}}].

\bibitem{DiDio:2018unb}
E.~Di~Dio, R.~Durrer, R.~Maartens, F.~Montanari, and O.~Umeh, {\it {The
  Full-Sky Angular Bispectrum in Redshift Space}},  {\em JCAP} {\bf 1904}
  (2019) 053, [\href{http://arxiv.org/abs/1812.09297}{{\tt arXiv:1812.09297}}].

\bibitem{Jelic-Cizmek:2020pkh}
G.~Jelic-Cizmek, F.~Lepori, C.~Bonvin, and R.~Durrer, {\it {On the importance
  of lensing for galaxy clustering in photometric and spectroscopic surveys}},
  {\em JCAP} {\bf 04} (2021) 055, [\href{http://arxiv.org/abs/2004.12981}{{\tt
  arXiv:2004.12981}}].

\bibitem{Cooray:2003hd}
A.~Cooray, D.~Huterer, and D.~Baumann, {\it {Growth rate of large scale
  structure as a powerful probe of dark energy}},  {\em Phys. Rev. D} {\bf 69}
  (2004) 027301, [\href{http://arxiv.org/abs/astro-ph/0304268}{{\tt
  astro-ph/0304268}}].

\bibitem{DiPorto:2007ovd}
C.~Di~Porto and L.~Amendola, {\it {Observational constraints on the linear
  fluctuation growth rate}},  {\em Phys. Rev. D} {\bf 77} (2008) 083508,
  [\href{http://arxiv.org/abs/0707.2686}{{\tt arXiv:0707.2686}}].

\bibitem{Tansella:2017rpi}
V.~Tansella, C.~Bonvin, R.~Durrer, B.~Ghosh, and E.~Sellentin, {\it {The
  full-sky relativistic correlation function and power spectrum of galaxy
  number counts. Part I: theoretical aspects}},  {\em JCAP} {\bf 1803} (2018),
  no.~03 019, [\href{http://arxiv.org/abs/1708.00492}{{\tt arXiv:1708.00492}}].

\bibitem{Tansella:2018sld}
V.~Tansella, G.~Jelic-Cizmek, C.~Bonvin, and R.~Durrer, {\it {COFFE: a code for
  the full-sky relativistic galaxy correlation function}},  {\em JCAP} {\bf
  1810} (2018), no.~10 032, [\href{http://arxiv.org/abs/1806.11090}{{\tt
  arXiv:1806.11090}}].

\bibitem{jeliccizmek:2020fl}
G.~Jelic-Cizmek, {\it The flat-sky approximation to galaxy number counts -
  redshift space correlation function},
  \href{http://arxiv.org/abs/2011.01878}{{\tt arXiv:2011.01878}}.

\end{thebibliography}\endgroup

\end{document}